\documentclass[]{aastex631}
\usepackage{amsmath}
\usepackage{float}
\usepackage{graphicx}
\usepackage{subfigure}
\usepackage{txfonts}
\usepackage{indentfirst}
\usepackage{ulem}
\usepackage[Symbolsmallscale]{upgreek}
\usepackage{nomencl}
\usepackage{enumitem}

\newcommand{\eq}[1]{\begin{equation}  #1 \end{equation}}
\newcommand{\eqs}[1]{\begin{equation} \begin{split} #1 \end{split} \end{equation}}

\newcommand{\br}[1]{\left( #1 \right)}

\newcommand{\bb}[1]{\left[ #1 \right]}

\newcommand{\dd}{{\rm d}}
\newcommand{\expo}[1]{~{\rm e}^{ #1 }}
\newcommand{\vek}[1]{\mbox{\boldmath $#1$}}

\newcommand{\ic}{{\rm i}}

\def\araa{ARA\&A}
\def\apj{ApJ}
\def\apjl{ApJ}

\def\aap{A\&A}

\def\mnras{MNRAS}
\def\nat{Nature}

\def\prd{Phys.~Rev.~D}

\begin{document}
\title{Lensing point-spread function of coherent astrophysical sources and non-trivial wave effects}
 
\author[0000-0003-2076-4510]{Xun Shi}
\email{xun@ynu.edu.cn}
\affiliation{South-Western Institute for Astronomy Research (SWIFAR), \\
Yunnan University, 650500 \\
Kunming, P. R. China}

\begin{abstract} 
    Most research on astrophysical lensing has been conducted using the geometric optics framework, where there exists a clear concept of lensing images.
    However, wave optics effects can be important for coherent sources, e.g. pulsars, fast raio bursts, and gravitational waves observed at long wavelengths.
    There, the concept of lensing images needs an extension.
    We introduce the concept of the `lensing point-spread function' (LPSF), the smoothed flux density distribution of a coherent point source after being lensed, as a generalization of the lensing image concept at finite frequencies.
    The frequency-dependent LPSF captures the gradual change of the flux density distribution of the source from discrete geometric images at high frequencies to a smooth distribution at low frequencies.
    It complements other generalizations of lensing images, notably the imaginary images and the Lefschetz thimbles.
    Being a footprint of a lensing system, the LPSF is useful for theoretical studies of lensing.
    Using the LPSF, we identify a frequency range with non-trivial wave effects, where both geometric optics and perturbative wave optics fail, and determine this range to be $|\kappa|^{-1} \lesssim \nu \lesssim 10$, with $\kappa$ and $\nu$ being the dimensionless lens amplitude and the reduced observing frequency, respectively.
    Observation of LPSFs with non-trivial wave effects requires either very close-by lenses or very large observing wavelengths.  
    The potential possibilities are the lensing of gravitational waves, the plasma lensing of Milky Way pulsars, and lensing by the solar gravitational lens.
\end{abstract}


\section{Introduction}
Astrophysical lensing is usually studied in the high-frequency, geometric limit using ray optics. 
The consideration of wave optics effects is traditionally limited to regions near caustics where ray optics fails \citep[e.g.][]{berry80} and to the study of one particular area of study: radio wave scintillation, e.g. that of pulsars in the interstellar medium \citep[e.g.][]{rickett90, stinebring01,walker04,cordes06,brisken10}.
There are mainly two reasons for wave optics not being widely used, despite that it offers a more complete description of lensing.
First, for most traditionally studied lensing systems, e.g. the lensing of galaxies and quasi-stellar objects, the Fresnel scale is significantly smaller in comparison to the sizes of the lenses. 
Thus, for these systems, wave optics effects can be safely ignored away from the caustics \citep{schneider92}.  
Second, it is intrinsically difficult to treat lensing with wave optics, since it involves integrating the highly oscillatory Fresnel-Kirchhoff diffraction integral, which is mathematically challenging.  
This limits the traditional considerations of wave optics to either perturbative expansions at the low-frequency limit, where the wave effects appear as small oscillatory modulations, or the eikonal approximation at the high-frequency limit where the wave effects are reduced to interference among the ray-optics-derived images.

Both these reasons have been invalidated by the developments over recent years.
New coherent sources on cosmological scales -- fast raio bursts (FRBs, \citealt{lorimer07, thornton13}) and gravitational waves \citep{abbott16, agazie23, epta23,reardon23, xu23} have been discovered, and are typically observed at long wavelengths, where wave optics effects can be important. 
On the other hand, new methods of integrating the Fresnel-Kirchhoff diffraction formula have been developed \citep{feldbrugge19, feldbrugge23, feldbrugge20, tambalo23}. 
In particular, by applying the Picard-Lefschetz theory (see \citealt{witten10} for a detailed description) and analytically continuing the integrand into the complex plane, the integration can be performed along the `Lefschetz thimbles’ -- a sum of steepest-descent contours where the integrand is non-oscillatory and rapidly converging \citep{feldbrugge19,feldbrugge20,feldbrugge23}.
Owing to these developments, there has been a renewed interest in wave optics effects in astrophysical lensing \citep[see, e.g.][]{main18, jow21,jow23, shi21, jow22, jow24b, suvorov22, caliskan23, leung23, savastano23, shi24, shi24b}.

In this article, we discuss the concept of lensing images in the wave optics framework. 
In the high-frequency, geometric limit, the observed flux of a source can be isolated into a discrete set of images which we refer to as `geometric images'.
The geometric images of a lensing system are the extrema of its Fermat potential, or equivalently, stationary phase points.
Their locations can be computed by solving the lens equation, and the flux associated with each of them can be obtained by evaluating the determinant of the lens-mapping Jacobian \citep[see, e.g.][]{schneider92}. 
At a finite frequency, these discrete geometric images no longer offer a complete description of the observed flux density distribution of the source. 
We show that there is a well-defined spatially smoothed flux density distribution at low frequencies which can be regarded as a generalization of the concept of lensing images.  

We refer to this smoothed flux density distribution as the `lensing Point Spread Function' (LPSF) and contrast it to other generalizations of lensing images to low frequencies from recent developments, notably, 
pieces of Lefschetz thimbles \citep{feldbrugge19,feldbrugge23, jow21} and the imaginary images \citep{grillo18,jow21}. 
While the latter are both helpful generalizations, the Lefschetz thimbles are fully conceptual and not directly observable. 
The imaginary images correspond to the imaginary roots of the lens equation, and thus they are still geometric images although they become effective only at low frequencies \citep{shi24}.
In contrast, the LPSF is a concept based on the wave optics framework which is valid at all frequencies, and it is in principle observable.

The structure of the paper is as follows. 
In Section\;\ref{sec:lensing}, we review the wave optics formulation of lensing. 
In Section\;\ref{sec:visual},  we introduce the concept of the LPSF and demonstrate its evolution with frequency for a few simple lensing systems.
We also compare the LPSF to the locations of the geometric images which helps illustrate the role of the imaginary images.
We use the LPSF to determine the conditions for non-trivial wave effects and discuss its observability in Section\;\ref{sec:obs}, and conclude in Section\;\ref{sec:conclusion}.

\section{lensing in wave optics}
\label{sec:lensing}
\subsection{Kirchhoff-Fresnel diffraction integral formulation of lensing}
The wave amplitude received by an observer for a point source with a unit magnitude is given by the Kirchhoff-Fresnel diffraction integral \citep[e.g.][]{schneider92}:
\eq{
    E(\vek{\beta}) = \br{\frac{\nu}{2\uppi \ic}} \int \exp\bb{\ic \nu \phi} \dd^2 \vek{x} \,.
    \label{eq:E}
} 

This expression of the wave amplitude $E(\vek{\beta})$, also referred to as the `transmission factor' of a lensing configuration, describes the diffraction of light at an intermediate phase screen by summing over all possible wave paths connecting the source, the observer, and a point $\vek{x}$ on the phase screen according to the principle of Huygens.
Polarization is ignored in this formulation as it is based on a scalar diffraction theory.
In Eq.\;\ref{eq:E}, $\vek{\beta}$ is the location of the source line of sight on the phase screen. 
The coordinates $\vek{x}$ and $\vek{\beta}$ are made dimensionless by scaling with the size of the lens $a_{\rm lens}$ following the convention of \citet{shi21} and \citet{jow23}.  
The reduced frequency $\nu$ is the key parameter that determines the importance of the wave effects. 
It is the square of the lens size $a_{\rm lens}$ / Fresnel scale $r_{\rm F}$ ratio that combines the observing wavelength $\lambda$ and the lensing distances:
\eq{
    \nu  = \frac{2\pi (1+z_{\rm lens}) a_{\rm lens}^2}{\lambda \bar{D}} = \frac{a_{\rm lens}^2}{r^2_{\rm F}} \,, 
}
where $\bar{D} = D_{\rm l} D_{\rm ls} / D_{\rm s}$ is a combination of the distances to the lens $D_{\rm l}$, the source $D_{\rm s}$, and that between the source and lens $D_{\rm ls}$. 
On cosmological scales, these distances should be interpreted as angular diameter distances, and one needs to include the factor with the lens redshift $z_{\rm lens}$ to account for the redshift of the light.



The phase function in the diffraction integral (Eq.\;\ref{eq:E}) can be written as
\eq{
    \phi(\vek{x}; \vek{\beta}, \kappa) = \frac{(\vek{x}-\vek{\beta})^2}{2} - \kappa\psi(\vek{x}) \,.
}
This incorporates the geometrical phase delay and the potential of a lens with amplitude $\kappa$ and a shape described by a function $\psi(x)$. 
We have adopted the convention that a converging lens (e.g. a gravitational lens) has a negative $\kappa$ value. 
Positive $\kappa$ values correspond to diverging lenses, which can be used to mimic overdense plasma lenses \citep[e.g.][]{clegg98, pen12, er14, cordes17, er18, er19, wagner20, shi21}.

\begin{figure*}
    \centering
    \includegraphics[width=.9\textwidth]{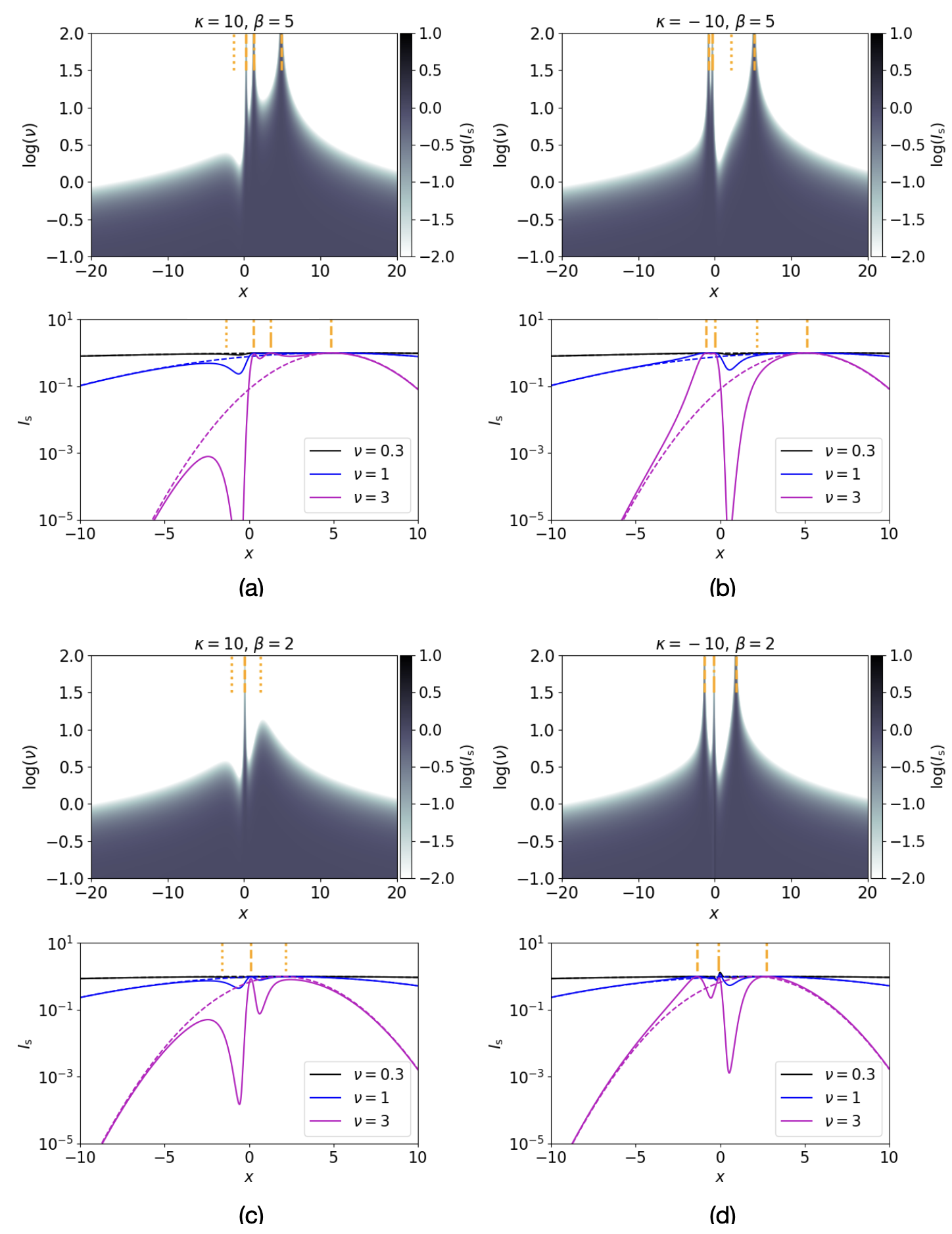} 
    \caption{Frequency dependence of the LPSF -- the smoothed flux density distribution $I_{\rm s}$ for 1D rational lenses. 
    We show both the $x-\nu$ distribution of $I_{\rm s}$ (the upper panel in each subplot) and the corresponding $I_{\rm s}(x)$ curves (lower panels) for various frequencies.
    The four subplots represent four lensing configurations, namely lenses with amplitudes $\kappa=\pm10$, with a plus sign for diverging lenses and a minus sign for converging lenses placed at two different source locations $\beta=5$ and $\beta=2$, respectively. 
    The flux density distribution is smoothed with a Gaussian kernel of width 0.1 along the $x-$axis. 
    We label the locations of the images in the geometric limit with the orange vertical lines, 
    and distinguish between real images with positive parity (dashed line), real images with negative parity (dash-dotted line), and effective imaginary images (dotted line).    
    The four subplots illustrate four different situations: (a) three real images and one observable imaginary image; (b) three real images and one indistinguishable imaginary image;
    (c) one real image and two imaginary images; (d) three real images and no imaginary image.
    In the $I_{\rm s}(x)$ plots, the dashed curves show the unlensed LPSFs given by Eq.\;\ref{eq:Is_unlensed}.
    }
    \label{fig:image_nu_kb} 
\end{figure*}

\section{LPSF of coherent sources}
\label{sec:visual}
\begin{figure*}
    \centering
    \includegraphics[width=1\textwidth]{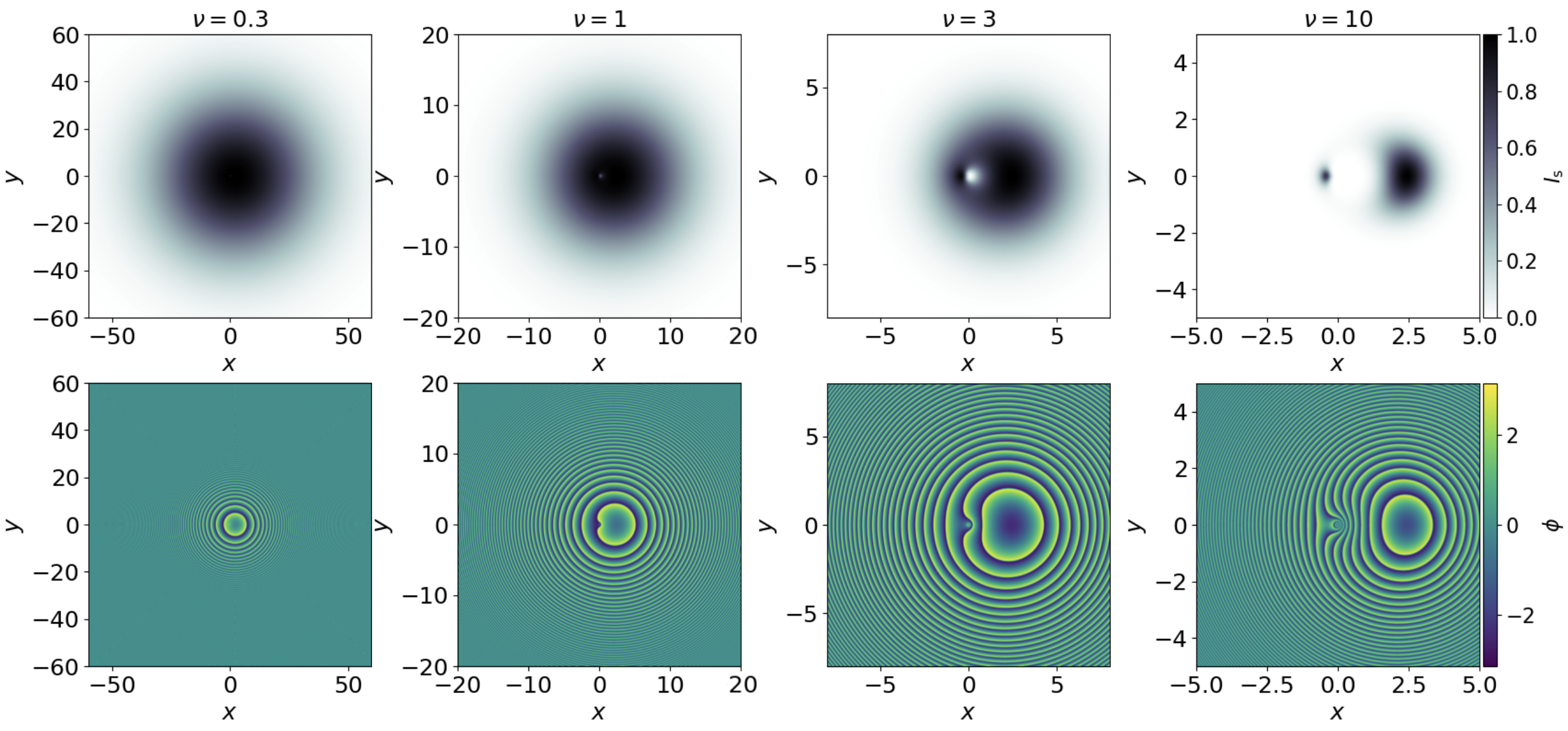} 
        \caption{
            2D LPSFs at different frequencies $\nu = 0.3, 1, 3,$ and 10 are shown in the upper panels for a converging point-mass lens $\psi=-\ln(\sqrt{x^2+y^2})$  with $\kappa=-1$ at $\beta=2$ displaced in the $x-$direction. 
            For this lensing configuration, there exist two real images at $y=0$, $x=2.4$ and $-0.4$, respectively. 
            The unit of the coordinates is the characteristic width of the lens, and the smoothing kernel is a 2D Gaussian of size $r_{\rm s} = 0.1$.
            The envelope of the diffuse brightness distribution can be well fitted by the 2D generalization of Eq.\;\ref{eq:Is_unlensed}: $I_{\rm s} \propto \exp\br{-\nu^2 r_{\rm s}^2 \br{(x-\beta)^2+y^2}}$, irrespective of the $\kappa$ value. 
            The corresponding phase function structures are shown in the lower panels.
            Note the different scales of the axes for the four columns.
        } 
        \label{fig:image2d} 
\end{figure*}

\begin{figure*}
    \centering
    \includegraphics[width=1\textwidth]{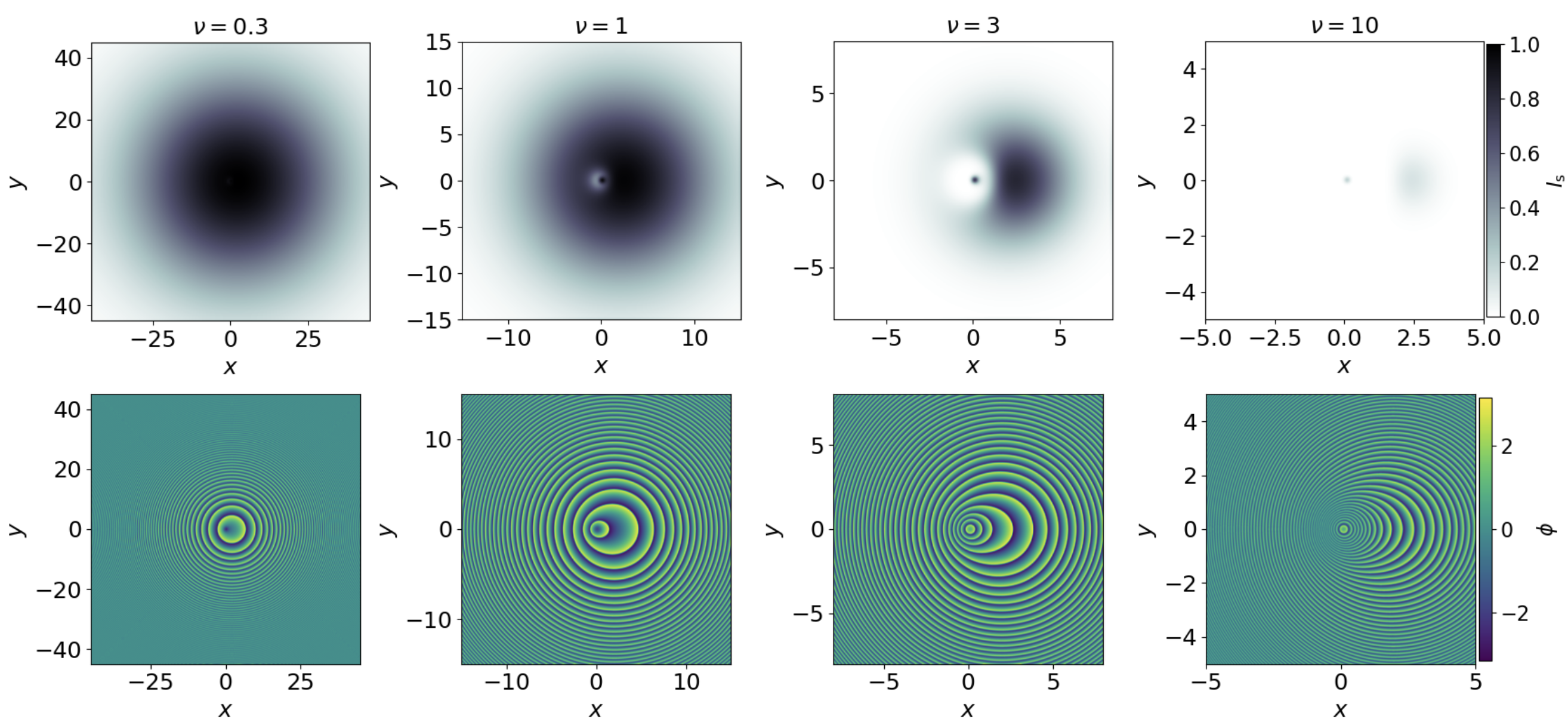} 
        \caption{
            Same as Fig.\;\ref{fig:image2d}, but for a diverging rational lens that mimics an overdense plasma lens.
            This axisymmetric lens with $\kappa=10$, $\beta=2$ is a generalization of the lens in panel (c) of Fig.\;\ref{fig:image_nu_kb} to 2D.  
            It creates the same images i.e. one real image and two imaginary images along the complexified x-axis. 
            Note that the two discrete patches of the LPSF at $\nu=10$ (upper right panel) include one corresponding to an imaginary image (the one to the right).
            }           
        \label{fig:image2d2} 
\end{figure*}

\subsection{Theory}
The concept of images in the geometric limit can be interpreted in two ways: 
(1) they are discrete copies of the source; and (2) they are the flux density distribution of the source in the sky as we observe it.

While the imaginary images and the Lefschetz thimbles (see Appendix.\;\ref{sec:other}) are generalizations of the 1st interpretation, 
what we would like to introduce here is a generalization of the 2nd interpretation, 
i.e., a description of the flux density distribution on the image plane at low frequencies. 
We shall refer to the image concept according to the 2nd interpretation as the LPSF to make a distinction.
The name derives from the fact that the LPSF describes how a coherent point source is spread out by lensing on the image plane.

The flux density of a coherent source is tricky to define \citep[see, e.g.][]{kim86}, 
and the same applies to the LPSF of coherent sources. 
To our knowledge, there have been limited previous efforts to obtain a flux density distribution for a lensing system considering wave optics, and they all follow a scheme initiated by \citet{nambu13}. 
They consider the image formation by a convex lens at the observer plane which yields a real image of the flux density distribution on the focal plane of the convex lens, whose form is given by a Fourier transform of the transmission factor $E(\vek{\beta})$ (Eq.\;\ref{eq:E}).

We, instead, do not assume any particular lens in the receiver system but base our LPSF definition just on the integrand of the diffraction integral, 
which is proportional to the wave amplitude on the image plane.
For a point source, this wave amplitude is 
\eq{
    U(\vek{x}) = \expo{\ic \nu \phi} \,.
}
It is the complex amplitude of disturbance contributed by point $\vek{x}$ on the phase screen.
The wave amplitude $U(\vek{x}; \beta)$ is a highly oscillatory function of $\vek{x}$.
However, an observing process inevitably introduces a smoothing of $U(\vek{x})$, 
and after a reasonable smoothing with a kernel $W(\vek{x})$, one obtains a smoothed wave field: 
\eq{
    U_{\rm s}(\vek{x}) = U(\vek{x}) \ast W(\vek{x}) \,.
}
This wavefield is equivalent to the electric field distribution on the focal plane of an additional lens as considered by \citet{nambu13}, 
with a smoothing introduced by the finite aperture of the lens.
We define the LPSF as the square intensity of this smoothed wave field, 
\eq{
    I_{\rm s}(\vek{x}) = |U_{\rm s}(\vek{x})|^2 \,.
}
It also describes the distribution of transverse momentum of photons coming from a point source through a lens.

The total flux is defined as a weighted mean of $I_{\rm s}(\vek{x})$ over the whole sky,
\eq{
    I = \frac{\nu}{2\uppi} \int I_{\rm s} \dd \vek{x} \,.
}
Clearly, both $I_{\rm s}(\vek{x})$ and $I$ depend on the smoothing.

An instructive example of the LPSF for an unlensed field smoothed with a Gaussian kernel is provided in appendix\;\ref{sec:unlensed}.
We shall limit ourselves to the case of point sources to highlight the wave effects.

\subsection{Results}
Fig.\;\ref{fig:image_nu_kb} shows a few 1D LPSFs at different frequencies.
These 1D examples highlight how the LPSF progressively changes from a few discrete geometric images at high frequencies to a smooth flux density distribution at low frequencies.
Whereas the LPSF at the high-frequency limit is well described by geometric lensing, at the low-frequency limit it is dominated by the smoothed unlensed field whose size is characterized by a combination of the Fresnel scale and the smoothing scale, $r_{\rm F}^2/r_{\rm s}$ (appendix\;\ref{sec:unlensed}).
This demonstrates the LPSF as a natural generalization of the concept of lensing images at low frequencies.

Especially noteworthy is the imprint of the imaginary images in the LPSF.
We mark the locations of the effective imaginary images with the dotted lines in Fig.\;\ref{fig:image_nu_kb} (see appendix.\;\ref{sec:other} for the definition of the locations of the imaginary images).
As expected, they do not show up in the LPSF at the high-frequency limit but only play a role at lower frequencies.
They form a peak in the LPSF at intermediate frequencies in some lensing configurations such as panels (a) and (c), but not in others like panel (b).
For an effective imaginary image created at a caustic like the one at $x=2.18$ in (c), the corresponding flux density peak is usually more pronounced.
In fact, as such an imaginary image can lie infinitely close to the real axis as the source location approaches the caustic, the flux density peak can also extend to high frequencies. 
On the other hand, effective imaginary images created at Stokes lines \citep{jow21,shi24},
such as the one in panel (b) do not usually create flux density peaks, but only modify the detailed shape of the LPSF at intermediate frequencies.
Apart from how far the imaginary image lies from the real axis, other factors can play a role in determining whether it can create a peak in the LPSF or not, 
including whether the imaginary image lies between real images and how much it is spatially separated from real images.

Fig.\;\ref{fig:image2d} and \ref{fig:image2d2} present 2D LPSFs and their corresponding phase function structures.
The converging point-mass lens in Fig.\;\ref{fig:image2d} is used to mimic gravitational lensing by compact objects, and the diverging lens in Fig.\;\ref{fig:image2d2} mimics plasma lensing by an overdense plasma concentration.
The basic picture presented in the 1D case is also valid in 2D.
At low frequencies, the LPSF is dominated by the smoothed unlensed field, and the flux density distribution is well-described by a Gaussian function apart from a tiny region around the lens.
As the frequency increases, the LPSF becomes more structured as the effect of the lens becomes more prominent. 

At high frequencies (e.g. $\nu=10$ for the lensing configurations in Figs.\;\ref{fig:image2d} and \ref{fig:image2d2}), the LPSFs become discrete patches around the geometric lensing images.
For the point-mass lens in Fig.\;\ref{fig:image2d}, the geometric images are two real images at $x=2.4, y=0$ and $x=-0.4, y=0$, respectively.
Note the LPSF morphology for this case: although it resembles the geometric lensing image for a resolved extended source exhibiting an Einstein ring, our source here is point-like and the diffuse flux density distribution is from the wave effects.
For the diverging rational lens in Fig.\;\ref{fig:image2d2}, the geometric images are one real image at $x=0.1, y=0$ and two imaginary images at $x=-1.6, y=0$ and $x=2.2, y=0$, respectively.
Unlike the real images which are extrema of the phase functions (lower panels in Figs.\;\ref{fig:image2d} and \ref{fig:image2d2}), the imaginary images do not appear as extrema in the $x-y$ space, but are only extrema in the complexified space.
Nevertheless, as in the 1D case, some imaginary images are capable of creating flux density peaks in the LPSF, e.g. the one around $x=2.2, y=0$ in the upper right panel of Fig.\;\ref{fig:image2d2}. 
These imaginary images are in principle observable.

We would like to promote the use of the phase function and the LPSF for theoretical studies of lensing systems.
Both of them are fingerprints of the lensing system that can be computed in an extremely straightforward way -- 
one needs neither to perform the oscillatory diffraction integral nor to solve the lens equation.
They give us a much more complete picture of the lensing system than the geometric images.

Observationally, it is in principle possible to infer the phase function structure and properties of the lens from the LPSF.
By its definition, the intensity of the LPSF directly reflects the spatial coherence of the phase function.

\begin{figure}
    \centering
    \includegraphics[width=0.6\textwidth]{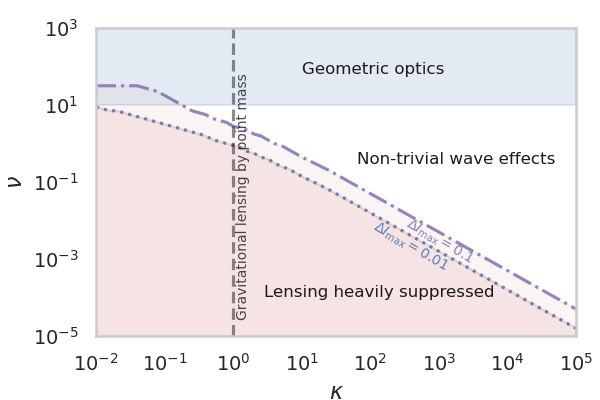} 
    \caption{Regimes of wave effects in lensing. 
    The parameter space is spanned by the lens amplitude $\kappa$ and the reduced frequency $\nu$. 
    Between the geometric optics regime at high frequencies $\nu \gtrsim 10$ (blue) and the low-frequency perturbative wave optics regime where lensing effects are heavily suppressed (red),
    there is a range of intermediate frequencies where non-trivial wave effects are observable (white).
    The lower limit of the regime of non-trivial wave effects can be  determined by the maximum difference $\Delta I_{\rm max}$ between the LPSF and the smoothed unlensed field.
    The dotted and dash-dotted lines show the limits defined using different threshold values of $\Delta I_{\rm max}$.
    They both follow $\kappa \nu = $ const. at large $\kappa$.   
    A 1D rational lens at $\beta=0$ and a smoothing kernel of width 0.1 are used for computing the LPSFs.
    }
    \label{fig:nu_wave}
\end{figure}

\section{Observability of non-trivial wave effects on the LPSF}
\label{sec:obs}

\subsection{Frequency range of non-trivial wave effects}
The LPSF is a continuation of the geometric images to low frequencies.
At the high-frequency limit, it reduces to the geometric images, and at extremely low frequencies, it approaches the smoothed unlensed field as lensing effects are heavily suppressed by diffraction.
Whereas geometric optics apply at the high-frequency limit, perturbative wave optics apply at the low-frequency limit.
Traditionally, the transition between the two regimes is defined by the Fresnel scale $r_{\rm F}$ in comparison to the lens size $a_{\rm lens}$, or equivalently, the reduced frequency $\nu$.
Wave effects are considered important when $\nu \lesssim 1$, and refraction (i.e. geometric optics) rather than diffraction is considered valid when $\nu \gg 1$ \citep{fiedler87, clegg98, dong18, kerr18}. 
In the field of pulsar scintillation where the lenses are complex, distributed density structures in the interstellar medium, the scattering of radio waves has been divided into refractive and diffractive regimes separated by $\nu$ i.e. the Fresnel scale criterion \citep{cordes86b, narayan92, johnson16, johnson18} when scattering is strong.
Associating the refractive regime with geometric optics and the diffractive regime with perturbative wave optics, 
the recent work of \citet{jow23} has raised a different opinion, stating that the combination of $|\kappa| \nu$ instead of $\nu$ itself determines the transition of the two regimes.

Using the LPSFs, it is straightforward to visualize these regimes for any given lensing configuration (see Fig.\;\ref{fig:image_nu_kb}).  
There is an intermediate range of frequencies where the LPSF is neither well-described by the geometric images nor by the smoothed unlensed field.
In this regime, both geometric optics and perturbative wave optics fail, and the wave effects are non-trivial.
The upper limit of this range of non-trivial wave effects is determined by where the LPSF starts to resemble the geometric images.
This is in turn determined by the size of the LPSF which is dependent solely on the reduced frequency $\nu$ (apart from the smoothing length). 
We take an upper limit $\nu_{\rm wave, max} \approx 10$ based on Fig.\;\ref{fig:image_nu_kb}.
The lower limit, on the other hand, is determined by the frequency at which the LPSF starts to deviate significantly from the unlensed field.
We compute this lower limit $\nu_{\rm wave, min}$ by identifying the reduced frequency where the maximum deviation of the LPSF from the unlensed field starts to be greater than a threshold value $\Delta I_{\rm max}$.
We find that $\nu_{\rm wave, min}$ depends strongly on the lens amplitude $|\kappa|$ in a similar way irrespective of the threshold value (Fig.\;\ref{fig:nu_wave}). 
At very large $|\kappa|$ values, $\nu_{\rm wave, min}$ is inversely proportional to $|\kappa|$.
Consequently, the onset of non-trivial wave effects at its low-frequency limit, i.e., where perturbative wave optics start to fail, is roughly set by $|\kappa| \nu \gtrsim 1$. 

Fig.\;\ref{fig:nu_wave} shows the $\kappa-\nu$ parameter space which can be divided into different regimes regarding the wave effects in lensing.
non-trivial wave effects exist in a triangular region well described by $|\kappa|^{-1} \lesssim \nu \lesssim 10$. 
The frequency range of non-trivial wave effects increases linearly with the lens amplitude $|\kappa|$.
For pulsar scintillation, the substructures on the secondary spectrum \citep[e.g.][]{hill05,brisken10,sprenger22} usually imply plasma lenses with very large amplitudes $\kappa \gg 1$ \citep{shi21b}. 
For these lenses, there is a large range of frequencies where non-trivial wave effects can be observable.
Note that we have treated the width $a_{\rm lens}$ and the amplitude $\kappa$ of the lens as independent variables. 
This is true for plasma lenses, but not entirely for gravitational lenses where both are related to the lens mass.
In the case of gravitational lensing by point-mass lenses, the $|\kappa|$ value is always unity (vertical dashed line in Fig.\;\ref{fig:nu_wave}), and the range of non-trivial wave effects reduces to the traditional Fresnel-scale criterion $1 \lesssim \nu \lesssim 10$.   
In the weak-lensing regime with $|\kappa| \ll 1$, no non-trivial wave effect can be observed.

\subsection{Observability}
Here we consider the observability of LPSFs with non-trivial wave effects.
For non-trivial wave effects to be observable, the reduced frequencies in the above-determined range must be physically reachable, and the LPSFs should also be resolvable.
Thus, we write the reduced frequency in terms of observable quantities $\theta_{\rm lens}$ (angular size of the lens), $\lambda$ (wavelength),  $D_{\rm l}$ (distance to the lens), and $D_{\rm ls}/D_{\rm s}$ (fractional distance of the lens to the source), as
\eqs{
    \nu 
    & = 4.4\times 10^6 (1+z_{\rm lens}) \br{\frac{\theta_{\rm lens}}{\rm arcsec}}^2 \br{\frac{D_{\rm l}D_{\rm s}/D_{\rm ls}}{\rm pc}} \br{\frac{\lambda}{\rm 1m}}^{-1} \,. \\
}
To bring $\nu$ into the range with non-trivial wave effects while keeping $\theta_{\rm lens}$ resolvable, one needs a very closeby lens and/or very large observing wavelength.

Gravitational waves, with their ultra-long wavelength, can easily fall into the range of non-trivial wave effects.
Among electromagnetic wave observations, one possible system is the plasma lensing of Milky Way pulsars. 
From typical values $\theta_{\rm lens} \sim 0.1$ mas, $\lambda \sim 0.1$m, $z_{\rm lens} = 0$, $D_{\rm l} \sim 100$ pc, and $D_{\rm ls}/D_{\rm s}=0.5$, we have $\nu \sim 0.9$ lying in the desired range. 
-- although to reach such a fine resolution of 0.1 mas, one needs a space-VLBI like the RadioAstron mission\footnote{http://www.asc.rssi.ru/radioastron/index.html}, or to use a scintillation screen as an interferometer \citep[see][]{pen14}.
Another rarely considered possibility is the solar gravitational lens \citep[e.g.][]{turyshev19,turyshev20, engeli22} which has been suggested for obtaining resolved images of nearby exoplanets and neutron stars.

\section{Conclusion}
\label{sec:conclusion}
In this study, we have introduced the concept of ‘LPSF’ as a way to generalize a traditional lensing image at finite frequencies. 
This new concept bridges the gap between the discrete geometric images observed at high frequencies and the smooth distributions seen at low frequencies, providing a more comprehensive framework for analyzing the frequency dependence of lensing phenomena.

The LPSF complements other recent generalizations like imaginary images and Lefschetz thimbles. 
Whereas the imaginary images and the Lefschetz thimbles maintain the discrete nature of geometric images, the LPSF generalizes another property of the geometric images -- it describes the flux density distribution of the source in the sky as we observe it at finite frequencies.
At high frequencies, the LPSF morphology is dominated by discrete geometric images including, in some lensing configurations, the imaginary images, implying that the latter are in principle observable.

Utilizing the LPSF, we have delineated the conditions under which non-trivial wave effects emerge, specifically within the range  $|\kappa|^{-1} \lesssim \nu \lesssim 10$, where both geometric optics and perturbative wave optics fail.
This has resolved the controversy about the transition between geometric optics and wave optics. 
The traditionally used Fresnel scale v.s. lens size comparison divides the geometric optics regime and the regime with non-trivial wave effects. 
On the other hand, the threshold defined by the combination of $|\kappa| \nu$ newly proposed by \citet{jow23} divides the regime with non-trivial wave effects and that of perturbative wave optics where lensing effects are heavily suppressed.

Apart from being useful for theoretical studies, the LPSFs are observable, even potentially in the regime of non-trivial wave effects. 
Resolving the LPSF in the regime of non-trivial wave effects requires very close-by lenses or very large observing wavelengths. 
Potential systems include the lensing of gravitational waves, the plasma lensing of Milky Way pulsars, and lensing by the solar gravitational lens.




\begin{acknowledgments}
    \noindent X.S. thanks the referee, Prasenjit Saha, for helpful comments and suggestions including the proposal of the more appropriate name ``lensing PSF'' for $I_s$. This work is supported by NSFC No. 12373025.
\end{acknowledgments}

\bibliographystyle{aasjournal}

\begin{thebibliography}{}
    \expandafter\ifx\csname natexlab\endcsname\relax\def\natexlab#1{#1}\fi
    \providecommand{\url}[1]{\href{#1}{#1}}
    \providecommand{\dodoi}[1]{doi:~\href{http://doi.org/#1}{\nolinkurl{#1}}}
    \providecommand{\doeprint}[1]{\href{http://ascl.net/#1}{\nolinkurl{http://ascl.net/#1}}}
    \providecommand{\doarXiv}[1]{\href{https://arxiv.org/abs/#1}{\nolinkurl{https://arxiv.org/abs/#1}}}
    
    \bibitem[{{Abbott} {et~al.}(2016){Abbott}, {Abbott}, {Abbott}, {Abernathy}, {Acernese}, {Ackley}, {Adams}, {Adams}, {Addesso}, {Adhikari}, {Adya}, {Affeldt}, {Agathos}, {Agatsuma}, {Aggarwal}, {Aguiar}, {Aiello}, {Ain}, {Ajith}, {Allen}, {Allocca}, {Altin}, {Anderson}, {Anderson}, {Arai}, {Arain}, {Araya}, {Arceneaux}, {Areeda}, {Arnaud}, {Arun}, {Ascenzi}, {Ashton}, {Ast}, {Aston}, {Astone}, {Aufmuth}, {Aulbert}, {Babak}, {Bacon}, {Bader}, {Baker}, {Baldaccini}, {Ballardin}, {Ballmer}, {Barayoga}, {Barclay}, {Barish}, {Barker}, {Barone}, {Barr}, {Barsotti}, {Barsuglia}, {Barta}, {Bartlett}, {Barton}, {Bartos}, {Bassiri}, {Basti}, {Batch}, {Baune}, {Bavigadda}, {Bazzan}, {Behnke}, {Bejger}, {Belczynski}, {Bell}, {Bell}, {Berger}, {Bergman}, {Bergmann}, {Berry}, {Bersanetti}, {Bertolini}, {Betzwieser}, {Bhagwat}, {Bhandare}, {Bilenko}, {Billingsley}, {Birch}, {Birney}, {Birnholtz}, {Biscans}, {Bisht}, {Bitossi}, {Biwer}, {Bizouard}, {Blackburn}, {Blair}, {Blair}, {Blair}, {Bloemen}, {Bock}, {Bodiya}, {Boer}, {Bogaert}, {Bogan}, {Bohe}, {Bojtos}, {Bond}, {Bondu}, {Bonnand}, {Boom}, {Bork}, {Boschi}, {Bose}, {Bouffanais}, {Bozzi}, {Bradaschia}, {Brady}, {Braginsky}, {Branchesi}, {Brau}, {Briant}, {Brillet}, {Brinkmann}, {Brisson}, {Brockill}, {Brooks}, {Brown}, {Brown}, {Brown}, {Buchanan}, {Buikema}, {Bulik}, {Bulten}, {Buonanno}, {Buskulic}, {Buy}, {Byer}, {Cabero}, {Cadonati}, {Cagnoli}, {Cahillane}, {Bustillo}, {Callister}, {Calloni}, {Camp}, {Cannon}, {Cao}, {Capano}, {Capocasa}, {Carbognani}, {Caride}, {Casanueva Diaz}, {Casentini}, {Caudill}, {Cavagli{\`a}}, {Cavalier}, {Cavalieri}, {Cella}, {Cepeda}, {Baiardi}, {Cerretani}, {Cesarini}, {Chakraborty}, {Chalermsongsak}, {Chamberlin}, {Chan}, {Chao}, {Charlton}, {Chassande-Mottin}, {Chen}, {Chen}, {Cheng}, {Chincarini}, {Chiummo}, {Cho}, {Cho}, {Chow}, {Christensen}, {Chu}, {Chua}, {Chung}, {Ciani}, {Clara}, {Clark}, {Cleva}, {Coccia}, {Cohadon}, {Colla}, {Collette}, {Cominsky}, {Constancio}, {Conte}, {Conti}, {Cook}, {Corbitt}, {Cornish}, {Corsi}, {Cortese}, {Costa}, {Coughlin}, {Coughlin}, {Coulon}, {Countryman}, {Couvares}, {Cowan}, {Coward}, {Cowart}, {Coyne}, {Coyne}, {Craig}, {Creighton}, {Creighton}, {Cripe}, {Crowder}, {Cruise}, {Cumming}, {Cunningham}, {Cuoco}, {Dal Canton}, {Danilishin}, {D'Antonio}, {Danzmann}, {Darman}, {Da Silva Costa}, {Dattilo}, {Dave}, {Daveloza}, {Davier}, {Davies}, {Daw}, {Day}, {De}, {DeBra}, {Debreczeni}, {Degallaix}, {De Laurentis}, {Del{\'e}glise}, {Del Pozzo}, {Denker}, {Dent}, {Dereli}, {Dergachev}, {DeRosa}, {De Rosa}, {DeSalvo}, {Dhurandhar}, {D{\'\i}az}, {Di Fiore}, {Di Giovanni}, {Di Lieto}, {Di Pace}, {Di Palma}, {Di Virgilio}, {Dojcinoski}, {Dolique}, {Donovan}, {Dooley}, {Doravari}, {Douglas}, {Downes}, {Drago}, {Drever}, {Driggers}, {Du}, {Ducrot}, {Dwyer}, {Edo}, {Edwards}, {Effler}, {Eggenstein}, {Ehrens}, {Eichholz}, {Eikenberry}, {Engels}, {Essick}, {Etzel}, {Evans}, {Evans}, {Everett}, {Factourovich}, {Fafone}, {Fair}, {Fairhurst}, {Fan}, {Fang}, {Farinon}, {Farr}, {Farr}, {Favata}, {Fays}, {Fehrmann}, {Fejer}, {Feldbaum}, {Ferrante}, {Ferreira}, {Ferrini}, {Fidecaro}, {Finn}, {Fiori}, {Fiorucci}, {Fisher}, {Flaminio}, {Fletcher}, {Fong}, {Fournier}, {Franco}, {Frasca}, {Frasconi}, {Frede}, {Frei}, {Freise}, {Frey}, {Frey}, {Fricke}, {Fritschel}, {Frolov}, {Fulda}, {Fyffe}, {Gabbard}, {Gair}, {Gammaitoni}, {Gaonkar}, {Garufi}, {Gatto}, {Gaur}, {Gehrels}, {Gemme}, {Gendre}, {Genin}, {Gennai}, {George}, {Gergely}, {Germain}, {Ghosh}, {Ghosh}, {Ghosh}, {Giaime}, {Giardina}, {Giazotto}, {Gill}, {Glaefke}, {Gleason}, {Goetz}, {Goetz}, {Gondan}, {Gonz{\'a}lez}, {Castro}, {Gopakumar}, {Gordon}, {Gorodetsky}, {Gossan}, {Gosselin}, {Gouaty}, {Graef}, {Graff}, {Granata}, {Grant}, {Gras}, {Gray}, {Greco}, {Green}, {Greenhalgh}, {Groot}, {Grote}, {Grunewald}, {Guidi}, {Guo}, {Gupta}, {Gupta}, {Gushwa}, {Gustafson}, {Gustafson}, {Hacker}, {Hall}, {Hall}, {Hammond}, {Haney}, {Hanke}, {Hanks}, {Hanna}, {Hannam}, {Hanson}, {Hardwick}, {Harms}, {Harry}, {Harry}, {Hart}, {Hartman}, {Haster}, {Haughian}, {Healy}, {Heefner}, {Heidmann}, {Heintze}, {Heinzel}, {Heitmann}, {Hello}, {Hemming}, {Hendry}, {Heng}, {Hennig}, {Heptonstall}, {Heurs}, {Hild}, {Hoak}, {Hodge}, {Hofman}, {Hollitt}, {Holt}, {Holz}, {Hopkins}, {Hosken}, {Hough}, {Houston}, {Howell}, {Hu}, {Huang}, {Huerta}, {Huet}, {Hughey}, {Husa}, {Huttner}, {Huynh-Dinh}, {Idrisy}, {Indik}, {Ingram}, {Inta}, {Isa}, {Isac}, {Isi}, {Islas}, {Isogai}, {Iyer}, {Izumi}, {Jacobson}, {Jacqmin}, {Jang}, {Jani}, {Jaranowski}, {Jawahar}, {Jim{\'e}nez-Forteza}, {Johnson}, {Johnson-McDaniel}, {Jones}, {Jones}, {Jonker}, {Ju}, {Haris}, {Kalaghatgi}, {Kalogera}, {Kandhasamy}, {Kang}, {Kanner}, {Karki}, {Kasprzack}, {Katsavounidis}, {Katzman}, {Kaufer}, {Kaur}, {Kawabe}, {Kawazoe}, {K{\'e}f{\'e}lian}, {Kehl}, {Keitel}, {Kelley}, {Kells}, {Kennedy}, {Keppel}, {Key}, {Khalaidovski}, {Khalili}, {Khan}, {Khan}, {Khan}, {Khazanov}, {Kijbunchoo}, {Kim}, {Kim}, {Kim}, {Kim}, {Kim}, {Kim}, {King}, {King}, {Kinzel}, {Kissel}, {Kleybolte}, {Klimenko}, {Koehlenbeck}, {Kokeyama}, {Koley}, {Kondrashov}, {Kontos}, {Koranda}, {Korobko}, {Korth}, {Kowalska}, {Kozak}, {Kringel}, {Krishnan}, {Kr{\'o}lak}, {Krueger}, {Kuehn}, {Kumar}, {Kumar}, {Kuo}, {Kutynia}, {Kwee}, {Lackey}, {Landry}, {Lange}, {Lantz}, {Lasky}, {Lazzarini}, {Lazzaro}, {Leaci}, {Leavey}, {Lebigot}, {Lee}, {Lee}, {Lee}, {Lee}, {Lenon}, {Leonardi}, {Leong}, {Leroy}, {Letendre}, {Levin}, {Levine}, {Li}, {Libson}, {Littenberg}, {Lockerbie}, {Logue}, {Lombardi}, {London}, {Lord}, {Lorenzini}, {Loriette}, {Lormand}, {Losurdo}, {Lough}, {Lousto}, {Lovelace}, {L{\"u}ck}, {Lundgren}, {Luo}, {Lynch}, {Ma}, {MacDonald}, {Machenschalk}, {MacInnis}, {Macleod}, {Maga{\~n}a-Sandoval}, {Magee}, {Mageswaran}, {Majorana}, {Maksimovic}, {Malvezzi}, {Man}, {Mandel}, {Mandic}, {Mangano}, {Mansell}, {Manske}, {Mantovani}, {Marchesoni}, {Marion}, {M{\'a}rka}, {M{\'a}rka}, {Markosyan}, {Maros}, {Martelli}, {Martellini}, {Martin}, {Martin}, {Martynov}, {Marx}, {Mason}, {Masserot}, {Massinger}, {Masso-Reid}, {Matichard}, {Matone}, {Mavalvala}, {Mazumder}, {Mazzolo}, {McCarthy}, {McClelland}, {McCormick}, {McGuire}, {McIntyre}, {McIver}, {McManus}, {McWilliams}, {Meacher}, {Meadors}, {Meidam}, {Melatos}, {Mendell}, {Mendoza-Gandara}, {Mercer}, {Merilh}, {Merzougui}, {Meshkov}, {Messenger}, {Messick}, {Meyers}, {Mezzani}, {Miao}, {Michel}, {Middleton}, {Mikhailov}, {Milano}, {Miller}, {Millhouse}, {Minenkov}, {Ming}, {Mirshekari}, {Mishra}, {Mitra}, {Mitrofanov}, {Mitselmakher}, {Mittleman}, {Moggi}, {Mohan}, {Mohapatra}, {Montani}, {Moore}, {Moore}, {Moraru}, {Moreno}, {Morriss}, {Mossavi}, {Mours}, {Mow-Lowry}, {Mueller}, {Mueller}, {Muir}, {Mukherjee}, {Mukherjee}, {Mukherjee}, {Mukund}, {Mullavey}, {Munch}, {Murphy}, {Murray}, {Mytidis}, {Nardecchia}, {Naticchioni}, {Nayak}, {Necula}, {Nedkova}, {Nelemans}, {Neri}, {Neunzert}, {Newton}, {Nguyen}, {Nielsen}, {Nissanke}, {Nitz}, {Nocera}, {Nolting}, {Normandin}, {Nuttall}, {Oberling}, {Ochsner}, {O'Dell}, {Oelker}, {Ogin}, {Oh}, {Oh}, {Ohme}, {Oliver}, {Oppermann}, {Oram}, {O'Reilly}, {O'Shaughnessy}, {Ott}, {Ottaway}, {Ottens}, {Overmier}, {Owen}, {Pai}, {Pai}, {Palamos}, {Palashov}, {Palomba}, {Pal-Singh}, {Pan}, {Pan}, {Pankow}, {Pannarale}, {Pant}, {Paoletti}, {Paoli}, {Papa}, {Paris}, {Parker}, {Pascucci}, {Pasqualetti}, {Passaquieti}, {Passuello}, {Patricelli}, {Patrick}, {Pearlstone}, {Pedraza}, {Pedurand}, {Pekowsky}, {Pele}, {Penn}, {Perreca}, {Pfeiffer}, {Phelps}, {Piccinni}, {Pichot}, {Pickenpack}, {Piergiovanni}, {Pierro}, {Pillant}, {Pinard}, {Pinto}, {Pitkin}, {Poeld}, {Poggiani}, {Popolizio}, {Post}, {Powell}, {Prasad}, {Predoi}, {Premachandra}, {Prestegard}, {Price}, {Prijatelj}, {Principe}, {Privitera}, {Prix}, {Prodi}, {Prokhorov}, {Puncken}, {Punturo}, {Puppo}, {P{\"u}rrer}, {Qi}, {Qin}, {Quetschke}, {Quintero}, {Quitzow-James}, {Raab}, {Rabeling}, {Radkins}, {Raffai}, {Raja}, {Rakhmanov}, {Ramet}, {Rapagnani}, {Raymond}, {Razzano}, {Re}, {Read}, {Reed}, {Regimbau}, {Rei}, {Reid}, {Reitze}, {Rew}, {Reyes}, {Ricci}, {Riles}, {Robertson}, {Robie}, {Robinet}, {Rocchi}, {Rolland}, {Rollins}, {Roma}, {Romano}, {Romano}, {Romanov}, {Romie}, {Rosi{\'n}ska}, {Rowan}, {R{\"u}diger}, {Ruggi}, {Ryan}, {Sachdev}, {Sadecki}, {Sadeghian}, {Salconi}, {Saleem}, {Salemi}, {Samajdar}, {Sammut}, {Sampson}, {Sanchez}, {Sandberg}, {Sandeen}, {Sanders}, {Sanders}, {Sassolas}, {Sathyaprakash}, {Saulson}, {Sauter}, {Savage}, {Sawadsky}, {Schale}, {Schilling}, {Schmidt}, {Schmidt}, {Schnabel}, {Schofield}, {Sch{\"o}nbeck}, {Schreiber}, {Schuette}, {Schutz}, {Scott}, {Scott}, {Sellers}, {Sengupta}, {Sentenac}, {Sequino}, {Sergeev}, {Serna}, {Setyawati}, {Sevigny}, {Shaddock}, {Shaffer}, {Shah}, {Shahriar}, {Shaltev}, {Shao}, {Shapiro}, {Shawhan}, {Sheperd}, {Shoemaker}, {Shoemaker}, {Siellez}, {Siemens}, {Sigg}, {Silva}, {Simakov}, {Singer}, {Singer}, {Singh}, {Singh}, {Singhal}, {Sintes}, {Slagmolen}, {Smith}, {Smith}, {Smith}, {Smith}, {Son}, {Sorazu}, {Sorrentino}, {Souradeep}, {Srivastava}, {Staley}, {Steinke}, {Steinlechner}, {Steinlechner}, {Steinmeyer}, {Stephens}, {Stevenson}, {Stone}, {Strain}, {Straniero}, {Stratta}, {Strauss}, {Strigin}, {Sturani}, {Stuver}, {Summerscales}, {Sun}, {Sutton}, {Swinkels}, {Szczepa{\'n}czyk}, {Tacca}, {Talukder}, {Tanner}, {T{\'a}pai}, {Tarabrin}, {Taracchini}, {Taylor}, {Theeg}, {Thirugnanasambandam}, {Thomas}, {Thomas}, {Thomas}, {Thorne}, {Thorne}, {Thrane}, {Tiwari}, {Tiwari}, {Tokmakov}, {Tomlinson}, {Tonelli}, {Torres}, {Torrie}, {T{\"o}yr{\"a}}, {Travasso}, {Traylor}, {Trifir{\`o}}, {Tringali}, {Trozzo}, {Tse}, {Turconi}, {Tuyenbayev}, {Ugolini}, {Unnikrishnan}, {Urban}, {Usman}, {Vahlbruch}, {Vajente}, {Valdes}, {Vallisneri}, {van Bakel}, {van Beuzekom}, {van den Brand}, {Van Den Broeck}, {Vander-Hyde}, {van der Schaaf}, {van Heijningen}, {van Veggel}, {Vardaro}, {Vass}, {Vas{\'u}th}, {Vaulin}, {Vecchio}, {Vedovato}, {Veitch}, {Veitch}, {Venkateswara}, {Verkindt},
      {Vetrano}, {Vicer{\'e}}, {Vinciguerra}, {Vine}, {Vinet}, {Vitale}, {Vo}, {Vocca}, {Vorvick}, {Voss}, {Vousden}, {Vyatchanin}, {Wade}, {Wade}, {Wade}, {Waldman}, {Walker}, {Wallace}, {Walsh}, {Wang}, {Wang}, {Wang}, {Wang}, {Wang}, {Ward}, {Ward}, {Warner}, {Was}, {Weaver}, {Wei}, {Weinert}, {Weinstein}, {Weiss}, {Welborn}, {Wen}, {We{\ss}els}, {Westphal}, {Wette}, {Whelan}, {Whitcomb}, {White}, {Whiting}, {Wiesner}, {Wilkinson}, {Willems}, {Williams}, {Williams}, {Williamson}, {Willis}, {Willke}, {Wimmer}, {Winkelmann}, {Winkler}, {Wipf}, {Wiseman}, {Wittel}, {Woan}, {Worden}, {Wright}, {Wu}, {Yablon}, {Yakushin}, {Yam}, {Yamamoto}, {Yancey}, {Yap}, {Yu}, {Yvert}, {Zadro{\.Z}ny}, {Zangrando}, {Zanolin}, {Zendri}, {Zevin}, {Zhang}, {Zhang}, {Zhang}, {Zhang}, {Zhao}, {Zhou}, {Zhou}, {Zhu}, {Zucker}, {Zuraw}, {Zweizig}, {LIGO Scientific Collaboration}, \& {Virgo Collaboration}}]{abbott16}
    {Abbott}, B.~P., {Abbott}, R., {Abbott}, T.~D., {et~al.} 2016, \prl, 116, 061102, \dodoi{10.1103/PhysRevLett.116.061102}
    
    \bibitem[{{Agazie} {et~al.}(2023){Agazie}, {Anumarlapudi}, {Archibald}, {Arzoumanian}, {Baker}, {B{\'e}csy}, {Blecha}, {Brazier}, {Brook}, {Burke-Spolaor}, {Burnette}, {Case}, {Charisi}, {Chatterjee}, {Chatziioannou}, {Cheeseboro}, {Chen}, {Cohen}, {Cordes}, {Cornish}, {Crawford}, {Cromartie}, {Crowter}, {Cutler}, {Decesar}, {Degan}, {Demorest}, {Deng}, {Dolch}, {Drachler}, {Ellis}, {Ferrara}, {Fiore}, {Fonseca}, {Freedman}, {Garver-Daniels}, {Gentile}, {Gersbach}, {Glaser}, {Good}, {G{\"u}ltekin}, {Hazboun}, {Hourihane}, {Islo}, {Jennings}, {Johnson}, {Jones}, {Kaiser}, {Kaplan}, {Kelley}, {Kerr}, {Key}, {Klein}, {Laal}, {Lam}, {Lamb}, {Lazio}, {Lewandowska}, {Littenberg}, {Liu}, {Lommen}, {Lorimer}, {Luo}, {Lynch}, {Ma}, {Madison}, {Mattson}, {McEwen}, {McKee}, {McLaughlin}, {McMann}, {Meyers}, {Meyers}, {Mingarelli}, {Mitridate}, {Natarajan}, {Ng}, {Nice}, {Ocker}, {Olum}, {Pennucci}, {Perera}, {Petrov}, {Pol}, {Radovan}, {Ransom}, {Ray}, {Romano}, {Sardesai}, {Schmiedekamp}, {Schmiedekamp}, {Schmitz}, {Schult}, {Shapiro-Albert}, {Siemens}, {Simon}, {Siwek}, {Stairs}, {Stinebring}, {Stovall}, {Sun}, {Susobhanan}, {Swiggum}, {Taylor}, {Taylor}, {Turner}, {Unal}, {Vallisneri}, {van Haasteren}, {Vigeland}, {Wahl}, {Wang}, {Witt}, {Young}, \& {Nanograv Collaboration}}]{agazie23}
    {Agazie}, G., {Anumarlapudi}, A., {Archibald}, A.~M., {et~al.} 2023, \apjl, 951, L8, \dodoi{10.3847/2041-8213/acdac6}
    
    \bibitem[{{Berry} \& {Upstill}(1980)}]{berry80}
    {Berry}, M.~V., \& {Upstill}, C. 1980, Progess in Optics, 18, 257, \dodoi{10.1016/S0079-6638(08)70215-4}
    
    \bibitem[{{Brisken} {et~al.}(2010){Brisken}, {Macquart}, {Gao}, {Rickett}, {Coles}, {Deller}, {Tingay}, \& {West}}]{brisken10}
    {Brisken}, W.~F., {Macquart}, J.~P., {Gao}, J.~J., {et~al.} 2010, \apj, 708, 232, \dodoi{10.1088/0004-637X/708/1/232}
    
    \bibitem[{{{\c{C}}al{\i}{\c{s}}kan} {et~al.}(2023){{\c{C}}al{\i}{\c{s}}kan}, {Anil Kumar}, {Ji}, {Ezquiaga}, {Cotesta}, {Berti}, \& {Kamionkowski}}]{caliskan23}
    {{\c{C}}al{\i}{\c{s}}kan}, M., {Anil Kumar}, N., {Ji}, L., {et~al.} 2023, \prd, 108, 123543, \dodoi{10.1103/PhysRevD.108.123543}
    
    \bibitem[{{Clegg} {et~al.}(1998){Clegg}, {Fey}, \& {Lazio}}]{clegg98}
    {Clegg}, A.~W., {Fey}, A.~L., \& {Lazio}, T. J.~W. 1998, \apj, 496, 253, \dodoi{10.1086/305344}
    
    \bibitem[{{Connor}(1973)}]{conner73}
    {Connor}, J.~N.~L. 1973, Molecular Physics, 25, 181, \dodoi{10.1080/00268977300100181}
    
    \bibitem[{{Cordes} {et~al.}(1986){Cordes}, {Pidwerbetsky}, \& {Lovelace}}]{cordes86b}
    {Cordes}, J.~M., {Pidwerbetsky}, A., \& {Lovelace}, R.~V.~E. 1986, \apj, 310, 737, \dodoi{10.1086/164728}
    
    \bibitem[{{Cordes} {et~al.}(2006){Cordes}, {Rickett}, {Stinebring}, \& {Coles}}]{cordes06}
    {Cordes}, J.~M., {Rickett}, B.~J., {Stinebring}, D.~R., \& {Coles}, W.~A. 2006, \apj, 637, 346, \dodoi{10.1086/498332}
    
    \bibitem[{{Cordes} {et~al.}(2017){Cordes}, {Wasserman}, {Hessels}, {Lazio}, {Chatterjee}, \& {Wharton}}]{cordes17}
    {Cordes}, J.~M., {Wasserman}, I., {Hessels}, J.~W.~T., {et~al.} 2017, \apj, 842, 35, \dodoi{10.3847/1538-4357/aa74da}
    
    \bibitem[{{Dong} {et~al.}(2018){Dong}, {Petropoulou}, \& {Giannios}}]{dong18}
    {Dong}, L., {Petropoulou}, M., \& {Giannios}, D. 2018, \mnras, 481, 2685, \dodoi{10.1093/mnras/sty2427}
    
    \bibitem[{{Engeli} \& {Saha}(2022)}]{engeli22}
    {Engeli}, S., \& {Saha}, P. 2022, \mnras, 516, 4679, \dodoi{10.1093/mnras/stac2522}
    
    \bibitem[{{EPTA Collaboration} {et~al.}(2023){EPTA Collaboration}, {InPTA Collaboration}, {Antoniadis}, {Arumugam}, {Arumugam}, {Babak}, {Bagchi}, {Bak Nielsen}, {Bassa}, {Bathula}, {Berthereau}, {Bonetti}, {Bortolas}, {Brook}, {Burgay}, {Caballero}, {Chalumeau}, {Champion}, {Chanlaridis}, {Chen}, {Cognard}, {Dandapat}, {Deb}, {Desai}, {Desvignes}, {Dhanda-Batra}, {Dwivedi}, {Falxa}, {Ferdman}, {Franchini}, {Gair}, {Goncharov}, {Gopakumar}, {Graikou}, {Grie{\ss}meier}, {Guillemot}, {Guo}, {Gupta}, {Hisano}, {Hu}, {Iraci}, {Izquierdo-Villalba}, {Jang}, {Jawor}, {Janssen}, {Jessner}, {Joshi}, {Kareem}, {Karuppusamy}, {Keane}, {Keith}, {Kharbanda}, {Kikunaga}, {Kolhe}, {Kramer}, {Krishnakumar}, {Lackeos}, {Lee}, {Liu}, {Liu}, {Lyne}, {McKee}, {Maan}, {Main}, {Mickaliger}, {Ni{\c{t}}u}, {Nobleson}, {Paladi}, {Parthasarathy}, {Perera}, {Perrodin}, {Petiteau}, {Porayko}, {Possenti}, {Prabu}, {Quelquejay Leclere}, {Rana}, {Samajdar}, {Sanidas}, {Sesana}, {Shaifullah}, {Singha}, {Speri}, {Spiewak}, {Srivastava}, {Stappers}, {Surnis}, {Susarla}, {Susobhanan}, {Takahashi}, {Tarafdar}, {Theureau}, {Tiburzi}, {van der Wateren}, {Vecchio}, {Venkatraman Krishnan}, {Verbiest}, {Wang}, {Wang}, \& {Wu}}]{epta23}
    {EPTA Collaboration}, {InPTA Collaboration}, {Antoniadis}, J., {et~al.} 2023, \aap, 678, A50, \dodoi{10.1051/0004-6361/202346844}
    
    \bibitem[{{Er} \& {Mao}(2014)}]{er14}
    {Er}, X., \& {Mao}, S. 2014, \mnras, 437, 2180, \dodoi{10.1093/mnras/stt2043}
    
    \bibitem[{{Er} \& {Rogers}(2018)}]{er18}
    {Er}, X., \& {Rogers}, A. 2018, \mnras, 475, 867, \dodoi{10.1093/mnras/stx3290}
    
    \bibitem[{{Er} \& {Rogers}(2019)}]{er19}
    ---. 2019, \mnras, 488, 5651, \dodoi{10.1093/mnras/stz2073}
    
    \bibitem[{{Feldbrugge} {et~al.}(2019){Feldbrugge}, {Pen}, \& {Turok}}]{feldbrugge19}
    {Feldbrugge}, J., {Pen}, U.-L., \& {Turok}, N. 2019, arXiv:1909.04632, arXiv:1909.04632.
    \newblock \doarXiv{1909.04632}
    
    \bibitem[{{Feldbrugge} {et~al.}(2023){Feldbrugge}, {Pen}, \& {Turok}}]{feldbrugge23}
    ---. 2023, Annals of Physics, 451, 169255, \dodoi{10.1016/j.aop.2023.169255}
    
    \bibitem[{{Feldbrugge} \& {Turok}(2020)}]{feldbrugge20}
    {Feldbrugge}, J., \& {Turok}, N. 2020, arXiv:2008.01154, arXiv:2008.01154, \dodoi{10.48550/arXiv.2008.01154}
    
    \bibitem[{{Fiedler} {et~al.}(1987){Fiedler}, {Dennison}, {Johnston}, \& {Hewish}}]{fiedler87}
    {Fiedler}, R.~L., {Dennison}, B., {Johnston}, K.~J., \& {Hewish}, A. 1987, \nat, 326, 675, \dodoi{10.1038/326675a0}
    
    \bibitem[{{Grillo} \& {Cordes}(2018)}]{grillo18}
    {Grillo}, G., \& {Cordes}, J. 2018, arXiv:1810.09058, arXiv:1810.09058.
    \newblock \doarXiv{1810.09058}
    
    \bibitem[{{Hill} {et~al.}(2005){Hill}, {Stinebring}, {Asplund}, {Berwick}, {Everett}, \& {Hinkel}}]{hill05}
    {Hill}, A.~S., {Stinebring}, D.~R., {Asplund}, C.~T., {et~al.} 2005, \apjl, 619, L171, \dodoi{10.1086/428347}
    
    \bibitem[{{Johnson} \& {Narayan}(2016)}]{johnson16}
    {Johnson}, M.~D., \& {Narayan}, R. 2016, \apj, 826, 170, \dodoi{10.3847/0004-637X/826/2/170}
    
    \bibitem[{{Johnson} {et~al.}(2018){Johnson}, {Narayan}, {Psaltis}, {Blackburn}, {Kovalev}, {Gwinn}, {Zhao}, {Bower}, {Moran}, {Kino}, {Kramer}, {Akiyama}, {Dexter}, {Broderick}, \& {Sironi}}]{johnson18}
    {Johnson}, M.~D., {Narayan}, R., {Psaltis}, D., {et~al.} 2018, \apj, 865, 104, \dodoi{10.3847/1538-4357/aadcff}
    
    \bibitem[{{Jow} {et~al.}(2021){Jow}, {Lin}, {Tyhurst}, \& {Pen}}]{jow21}
    {Jow}, D.~L., {Lin}, F.~X., {Tyhurst}, E., \& {Pen}, U.-L. 2021, \mnras, 507, 5390, \dodoi{10.1093/mnras/stab2337}
    
    \bibitem[{{Jow} \& {Pen}(2022)}]{jow22}
    {Jow}, D.~L., \& {Pen}, U.-L. 2022, \mnras, 514, 4069, \dodoi{10.1093/mnras/stac1652}
    
    \bibitem[{{Jow} \& {Pen}(2024)}]{jow24b}
    ---. 2024, arXiv e-prints, arXiv:2407.03214, \dodoi{10.48550/arXiv.2407.03214}
    
    \bibitem[{{Jow} {et~al.}(2023){Jow}, {Pen}, \& {Feldbrugge}}]{jow23}
    {Jow}, D.~L., {Pen}, U.-L., \& {Feldbrugge}, J. 2023, \mnras, 525, 2107, \dodoi{10.1093/mnras/stad2332}
    
    \bibitem[{{Kerr} {et~al.}(2018){Kerr}, {Coles}, {Ward}, {Johnston}, {Tuntsov}, \& {Shannon}}]{kerr18}
    {Kerr}, M., {Coles}, W.~A., {Ward}, C.~A., {et~al.} 2018, \mnras, 474, 4637, \dodoi{10.1093/mnras/stx3101}
    
    \bibitem[{{Kim}(1986)}]{kim86}
    {Kim}, K.-J. 1986, Nuclear Instruments and Methods in Physics Research A, 246, 71, \dodoi{10.1016/0168-9002(86)90048-3}
    
    \bibitem[{{Leung} {et~al.}(2023){Leung}, {Jow}, {Saha}, {Dai}, {Oguri}, \& {Koopmans}}]{leung23}
    {Leung}, C., {Jow}, D., {Saha}, P., {et~al.} 2023, arXiv e-prints, arXiv:2304.01202, \dodoi{10.48550/arXiv.2304.01202}
    
    \bibitem[{{Lorimer} {et~al.}(2007){Lorimer}, {Bailes}, {McLaughlin}, {Narkevic}, \& {Crawford}}]{lorimer07}
    {Lorimer}, D.~R., {Bailes}, M., {McLaughlin}, M.~A., {Narkevic}, D.~J., \& {Crawford}, F. 2007, Science, 318, 777, \dodoi{10.1126/science.1147532}
    
    \bibitem[{{Main} {et~al.}(2018){Main}, {Yang}, {Chan}, {Li}, {Lin}, {Mahajan}, {Pen}, {Vanderlinde}, \& {van Kerkwijk}}]{main18}
    {Main}, R., {Yang}, I.~S., {Chan}, V., {et~al.} 2018, \nat, 557, 522, \dodoi{10.1038/s41586-018-0133-z}
    
    \bibitem[{{Nambu}(2013)}]{nambu13}
    {Nambu}, Y. 2013, International Journal of Astronomy and Astrophysics, 3, 1, \dodoi{10.4236/ijaa.2013.31001}
    
    \bibitem[{{Narayan}(1992)}]{narayan92}
    {Narayan}, R. 1992, Philosophical Transactions of the Royal Society of London Series A, 341, 151, \dodoi{10.1098/rsta.1992.0090}
    
    \bibitem[{{Pen} \& {King}(2012)}]{pen12}
    {Pen}, U.-L., \& {King}, L. 2012, \mnras, 421, L132, \dodoi{10.1111/j.1745-3933.2012.01223.x}
    
    \bibitem[{{Pen} {et~al.}(2014){Pen}, {Macquart}, {Deller}, \& {Brisken}}]{pen14}
    {Pen}, U.~L., {Macquart}, J.~P., {Deller}, A.~T., \& {Brisken}, W. 2014, \mnras, 440, L36, \dodoi{10.1093/mnrasl/slu010}
    
    \bibitem[{{Reardon} {et~al.}(2023){Reardon}, {Zic}, {Shannon}, {Hobbs}, {Bailes}, {Di Marco}, {Kapur}, {Rogers}, {Thrane}, {Askew}, {Bhat}, {Cameron}, {Cury{\l}o}, {Coles}, {Dai}, {Goncharov}, {Kerr}, {Kulkarni}, {Levin}, {Lower}, {Manchester}, {Mandow}, {Miles}, {Nathan}, {Os{\l}owski}, {Russell}, {Spiewak}, {Zhang}, \& {Zhu}}]{reardon23}
    {Reardon}, D.~J., {Zic}, A., {Shannon}, R.~M., {et~al.} 2023, \apjl, 951, L6, \dodoi{10.3847/2041-8213/acdd02}
    
    \bibitem[{{Rickett}(1990)}]{rickett90}
    {Rickett}, B.~J. 1990, \araa, 28, 561, \dodoi{10.1146/annurev.aa.28.090190.003021}
    
    \bibitem[{{Savastano} {et~al.}(2023){Savastano}, {Tambalo}, {Villarrubia-Rojo}, \& {Zumalac{\'a}rregui}}]{savastano23}
    {Savastano}, S., {Tambalo}, G., {Villarrubia-Rojo}, H., \& {Zumalac{\'a}rregui}, M. 2023, \prd, 108, 103532, \dodoi{10.1103/PhysRevD.108.103532}
    
    \bibitem[{{Schneider} {et~al.}(1992){Schneider}, {Ehlers}, \& {Falco}}]{schneider92}
    {Schneider}, P., {Ehlers}, J., \& {Falco}, E.~E. 1992, {Gravitational Lenses} (Berlin: Springer), \dodoi{10.1007/978-3-662-03758-4}
    
    \bibitem[{{Shi}(2021)}]{shi21b}
    {Shi}, X. 2021, \mnras, 508, 125, \dodoi{10.1093/mnras/stab2522}
    
    \bibitem[{{Shi}(2024{\natexlab{a}})}]{shi24}
    ---. 2024{\natexlab{a}}, \apj, 972, 118, \dodoi{10.3847/1538-4357/ad6761}
    
    \bibitem[{{Shi}(2024{\natexlab{b}})}]{shi24b}
    ---. 2024{\natexlab{b}}, \mnras, 534, 1143, \dodoi{10.1093/mnras/stae2127}
    
    \bibitem[{{Shi} \& {Xu}(2021)}]{shi21}
    {Shi}, X., \& {Xu}, Z. 2021, \mnras, 506, 6039, \dodoi{10.1093/mnras/stab2108}
    
    \bibitem[{{Sprenger} {et~al.}(2022){Sprenger}, {Main}, {Wucknitz}, {Mall}, \& {Wu}}]{sprenger22}
    {Sprenger}, T., {Main}, R., {Wucknitz}, O., {Mall}, G., \& {Wu}, J. 2022, \mnras, 515, 6198, \dodoi{10.1093/mnras/stac2160}
    
    \bibitem[{{Stinebring} {et~al.}(2001){Stinebring}, {McLaughlin}, {Cordes}, {Becker}, {Goodman}, {Kramer}, {Sheckard}, \& {Smith}}]{stinebring01}
    {Stinebring}, D.~R., {McLaughlin}, M.~A., {Cordes}, J.~M., {et~al.} 2001, \apjl, 549, L97, \dodoi{10.1086/319133}
    
    \bibitem[{{Suvorov}(2022)}]{suvorov22}
    {Suvorov}, A.~G. 2022, \apj, 930, 13, \dodoi{10.3847/1538-4357/ac5f45}
    
    \bibitem[{{Tambalo} {et~al.}(2023){Tambalo}, {Zumalac{\'a}rregui}, {Dai}, \& {Cheung}}]{tambalo23}
    {Tambalo}, G., {Zumalac{\'a}rregui}, M., {Dai}, L., \& {Cheung}, M. H.-Y. 2023, \prd, 108, 043527, \dodoi{10.1103/PhysRevD.108.043527}
    
    \bibitem[{{Thornton} {et~al.}(2013){Thornton}, {Stappers}, {Bailes}, {Barsdell}, {Bates}, {Bhat}, {Burgay}, {Burke-Spolaor}, {Champion}, {Coster}, {D'Amico}, {Jameson}, {Johnston}, {Keith}, {Kramer}, {Levin}, {Milia}, {Ng}, {Possenti}, \& {van Straten}}]{thornton13}
    {Thornton}, D., {Stappers}, B., {Bailes}, M., {et~al.} 2013, Science, 341, 53, \dodoi{10.1126/science.1236789}
    
    \bibitem[{{Turyshev} \& {Toth}(2019)}]{turyshev19}
    {Turyshev}, S.~G., \& {Toth}, V.~T. 2019, \prd, 100, 084018, \dodoi{10.1103/PhysRevD.100.084018}
    
    \bibitem[{{Turyshev} \& {Toth}(2020)}]{turyshev20}
    ---. 2020, \prd, 102, 024038, \dodoi{10.1103/PhysRevD.102.024038}
    
    \bibitem[{{Wagner} \& {Er}(2020)}]{wagner20}
    {Wagner}, J., \& {Er}, X. 2020, arXiv:2006.16263, arXiv:2006.16263.
    \newblock \doarXiv{2006.16263}
    
    \bibitem[{{Walker} {et~al.}(2004){Walker}, {Melrose}, {Stinebring}, \& {Zhang}}]{walker04}
    {Walker}, M.~A., {Melrose}, D.~B., {Stinebring}, D.~R., \& {Zhang}, C.~M. 2004, \mnras, 354, 43, \dodoi{10.1111/j.1365-2966.2004.08159.x}
    
    \bibitem[{{Witten}(2010)}]{witten10}
    {Witten}, E. 2010, arXiv:1009.6032, arXiv:1009.6032.
    \newblock \doarXiv{1009.6032}
    
    \bibitem[{{Xu} {et~al.}(2023){Xu}, {Chen}, {Guo}, {Jiang}, {Wang}, {Xu}, {Xue}, {Nicolas Caballero}, {Yuan}, {Xu}, {Wang}, {Hao}, {Luo}, {Lee}, {Han}, {Jiang}, {Shen}, {Wang}, {Wang}, {Xu}, {Wu}, {Manchester}, {Qian}, {Guan}, {Huang}, {Sun}, \& {Zhu}}]{xu23}
    {Xu}, H., {Chen}, S., {Guo}, Y., {et~al.} 2023, Research in Astronomy and Astrophysics, 23, 075024, \dodoi{10.1088/1674-4527/acdfa5}
    
    \end{thebibliography}

\appendix

\section{LPSF in an idealized case}
\label{sec:unlensed}
We consider the instructive example of the LPSF for a 1D unlensed point source smoothed by a Gaussian kernel. 
For this idealized case, the LPSF can be computed analytically. 

Taking a Gaussian kernel in a standardized form, 
\eq{
    G(x) = \frac{1}{\sqrt{2\uppi} r_{\rm s}}\expo{-\frac{x^2}{2 r_{\rm s}^2}} \,,
}    
the resulting smoothed field distribution $U_{\rm s}$ is still a Gaussian function, 
\eq{
    U_{\rm s}(x) = U(x)\ast G(x)= \sqrt{\frac{\ic}{(\nu r_{\rm s}^2 + \ic)}} \expo{-\frac{\nu^2 r_{\rm s}^2 - \ic \nu}{1+\nu^2 r_{\rm s}^4}\frac{(x-\beta)^2}{2}} \,,
}
and the LPSF is
\eq{
    I_{\rm s}(x) \equiv |U_{\rm s}|^2 = \frac{1}{\sqrt{1+\nu^2 r_{\rm s}^4}} \expo{-\frac{\nu^2 r_{\rm s}^2}{1+\nu^2 r_{\rm s}^4}(x-\beta)^2} \,.
    \label{eq:Is_unlensed}
}
Its form is a Gaussian with width $\sigma = \sqrt{(1+\nu^2 r_{\rm s}^4)/2}/\nu/r_{\rm s}$. 
In the regime where the smoothing kernel is much wider than the Fresnel scale, ${r_{\rm s}}/{r_{\rm F}} = \nu^{1/2} r_{\rm s} \gg 1$, 
the width of the LPSF is dominated by the smoothing kernel, $\sigma \approx r_{\rm s}/\sqrt{2}$, with no dependence on frequency.
In the opposite regime where  ${r_{\rm s}}/{r_{\rm F}} \ \ll 1$,  $\sigma \approx (\sqrt{2}\nu r_{\rm s})^{-1} = r_{\rm F}^2/(\sqrt{2}r_{\rm s})$, the width of the image depends on both the Fresnel scale and the smoothing scale.

The total flux $I \propto \sqrt{\frac{1}{\nu x_{\rm s}^2}} \propto r_{\rm F} / r_{\rm s}$, also dependent on both the Fresnel scale and the smoothing scale.

\section{Other generalizations of lensing images at finite frequencies}
\label{sec:other}
\subsection{Imaginary images}
At the high-frequency $\nu \to \infty$ limit, only the stationary points of the phase function contribute to the diffraction integral.
These stationary points can be obtained by solving the lens equation, which is derived from the stationary phase condition $\dd\phi/\dd\vek{x} = 0$.

Traditionally, it has been taken for granted that only the real solutions of the lens equation, referred to as the `geometric images', would contribute to the signal received by the observer. 
Nonetheless, the lens equation can yield complex solutions that have non-zero imaginary parts. 
It has been shown that these imaginary images can have a significant impact both near the caustics \citep{grillo18} and in weak-lensing regimes \citep{jow21, shi24}. 
They can and should be taken into account in the eikonal, or stationary phase approximation. 

The field associated with an imaginary image $i$ in the eikonal limit is \citep{conner73, grillo18}
\eq{
    E_i = \frac{\expo{\ic \nu \phi_i}}{\Delta_i^{1/2}}
}
with subcaption $i$ indicating evaluation at the location of image $i$, and
\eq{
    \Delta = \rm{det} \frac{\partial^2 \phi}{\partial \vek{x}^2} \,.
}
The integrand has been generalized to the complex domain. 
Correspondingly, the phase function has both real and imaginary parts:
\eq{
    \ic \phi \equiv h + \ic H \,.
}

The magnifications of geometric images are \citep{schneider92}
\eq{
 \mu_i \equiv |E_i|^2 = |\Delta_i|^{-1}.
}
This can be generalized to imaginary images as   
\eq{
 \mu_i = |\Delta_i|^{-1} \expo{2 \nu h_i}
}
where $h_i$ is the real part of the phase $\ic \phi_i$, with $h_i < 0$ for an imaginary image and $h_i = 0$ for a real image.  

\subsection{Lefschetz thimbles}
\begin{figure*}
    \centering
    \includegraphics[width=0.65\textwidth]{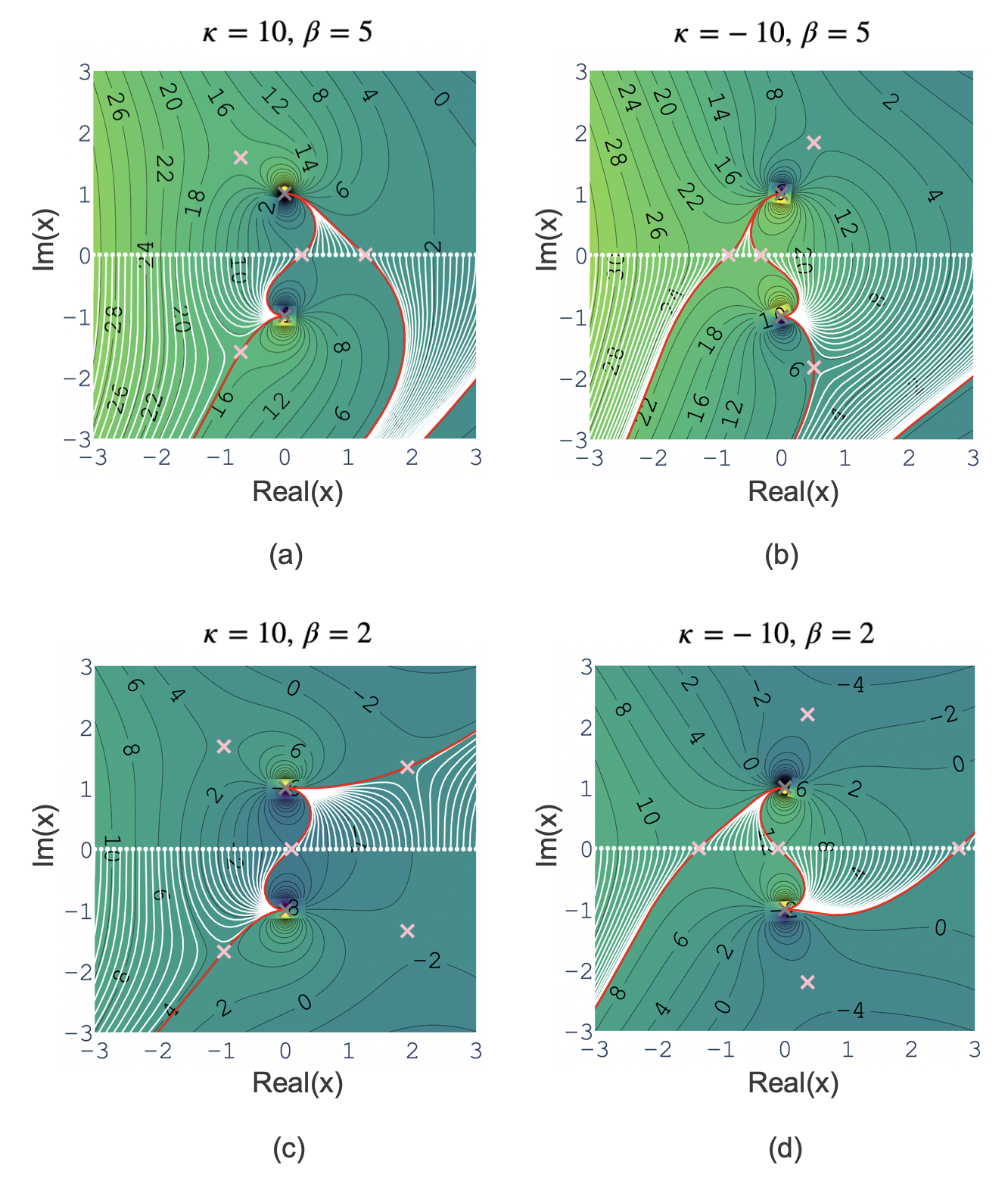}     
        \caption{
            Lefschetz thimbles (red) and flow lines (white) for 1D rational lenses $\psi = 1/(1+x^2)$. 
            The images are marked with pink crosses, and the poles are marked with gray crosses. 
            Contours of the $H$ function are shown in the background.
            The four subplots correspond to the four lensing configurations in Fig.\;\ref{fig:image_nu_kb}.
        } 
        \label{fig:thimble} 
\end{figure*}
The Picard-Lefschetz theory has recently been introduced to evaluate diffraction integrals in lensing by \citet{feldbrugge19, feldbrugge23}.  
By generalizing the integrand to the complex plane,
the Picard-Lefschetz theory deforms the integration contour into a sum of steepest-descent contours of the $h$ function in the complex plane. 
The integration can then be performed along these `Lefschetz thimbles' where the integrand is non-oscillatory (where the $H$ function is piecewise constant) and rapidly converging.

Fig.\;\ref{fig:thimble} presents the Lefschetz thimbles for the lensing systems in Fig.\;\ref{fig:image2d}.
The thimbles are shown as the red contours, and the flow lines \citep{shi24b} are shown as the white contours. 
They are pathlines of the downward flow of the $h$ function for points on the real axis.
The Lefschetz thimbles are a collection of smooth sub-contours, each corresponding to a specific image (pink crosses in Fig.\;\ref{fig:thimble}). 
In addition to the Lefschetz thimbles associated with geometric images, there will generally be contributions from imaginary images. 
Note, however, that not all imaginary images are associated with a Lefschetz thimble.
The imaginary images associated with Lefschetz thimbles are the effective images that are connected to flow lines whose origins on the real axis can be regarded as their real locations.
They contribute to the observed flux at finite observing frequencies.
Both the imaginary images and the Lefschetz thimbles maintain the notion of a finite collection of images even at low frequencies when geometric optics fail.

\end{document}